\documentclass{agujournal}
\journalname{Space Weather}

\begin{document}

\title{The importance of ensemble techniques for operational space weather forecasting}

\authors{1}{Sophie A. Murray}
\affiliation{1}{Trinity College Dublin, Ireland}
\correspondingauthor{Sophie A. Murray}{sophie.murray@tcd.ie}


\begin{keypoints}
\item Ensemble modeling methods are well-established in terrestrial weather forecasting and data science.
\item The application of ensemble methods to existing forecasts has improved their accuracy.
\item Ensemble techniques enable quantification of uncertainty in space weather predictions.
\end{keypoints}


\begin{abstract}
The space weather community has begun to use frontier methods such as data assimilation, machine learning, and ensemble modeling to advance current operational forecasting efforts. This was highlighted by a multi-disciplinary session at the 2017 American Geophysical Union Meeting, `Frontier Solar-Terrestrial Science Enabled by the Combination of Data-Driven Techniques and Physics-Based Understanding', with considerable discussion surrounding ensemble techniques. Here ensemble methods are described in detail; using a set of predictions to improve on a single-model output, for example taking a simple average of multiple models, or using more complex techniques for data assimilation. They have been used extensively in fields such as numerical weather prediction and data science, for both improving model accuracy and providing a measure of model uncertainty. Researchers in the space weather community have found them to be similarly useful, and some examples of success stories are highlighted in this commentary. Future developments are also encouraged to transition these basic research efforts to operational forecasting as well as providing prediction errors to aid end-user understanding.
\end{abstract}

\noindent \textbf{Plain Language Summary}
\

\noindent Space weather has the potential to impact a range of vital technologies and infrastructures that society has become dependent on in recent decades. Space weather forecasts receive the close attention of private companies, government policy makers, and even the general public hoping to spot aurorae. To that end there is an increasing need to improve the accuracy of forecasts considering they currently do not perform as well compared to weather forecasts. The space weather community has begun to try using methods used by more established communities to aid these efforts, such as in meteorology and data science. Ensembles were a topic of interesting discussion at a 2017 American Geophysical Union session dedicated to presenting these frontier methods, a technique which combines multiple models to produce a more accurate result. They have thus far proven extremely useful to improve model accuracy, and also can easily provide an error with the issued forecast, giving end-users a clearer picture of the real value of a forecast. This commentary encourages the continuing use of ensembles in space weather research, and to transition these efforts to advance operational space weather forecasting.


\section{Background}

Space weather has begun to receive the close attention of industry, government, and even the media in recent years, which is unsurprising considering the potentially damaging consequences a Carrington-level solar eruptive event could have on the technologies and infrastructures on which society has become increasingly dependent. However space weather forecasting still lags behind its terrestrial counterpart \citep{koskinen17}, and there is a need to improve the accuracy of current operational forecasts since many still barely outperform climatology \citep[see e.g.,][for a solar flare forecasting example]{barnes16}.

There has been a substantial push within the space weather community to try new techniques used by other research and operational communities to enhance current predictions. These frontier techniques have been helpful for the advancement of space weather forecasting efforts, for example borrowing forecasting and verification tools from terrestrial weather forecasters \citep{henley17, sharpe17}, or trying out new machine learning techniques used by data scientists \citep{mcgranaghan17}. This topic was explored in depth at the American Geophysical Union 2017 Fall Meeting during the multi-disciplinary session, `Frontier Solar-Terrestrial Science Enabled by the Combination of Data-Driven Techniques and Physics-Based Understanding' (\hbox{https://agu.confex.com/agu/fm17/meetingapp.cgi/Session/32931}). 

Organised by R. McGranaghan, J. Bortnik, E. Camporeale, and T. Matsuo, the session included talks on topics such as causality \citep{johnson17} and machine learning for solar eruptive forecasting \citep{bobra17}. Ensemble forecasts were an overarching theme which inspired interesting discussion throughout the session, with multiple talks on ensemble modeling for various different aspects of the space weather system \citep{dikpati17, morley17, murray17a}. As alluded to by \citet{henley17}, ensembles are a staple of weather forecasts worldwide, and are rapidly becoming a go-to method for improving the accuracy of space weather models \citep{knipp16}.


\section{Ensemble Modeling}

Instead of using a single prediction, ensemble methods use a set of predictions. For example, a common application is the multi-model ensemble, which combines different models to improve on a single-model prediction. Another approach uses a single model, but perturbs its initial conditions or parameter settings or schemes in order to produce multiple results. Varying input data, which in some cases are used as boundary conditions, can also be used to create an ensemble. More complex approaches are used with data assimilation techniques, where observations are inserted into models to nudge them closer to real conditions. For example, the popular Ensemble Kalman Filter (EnKF) is a Monte Carlo approximation of the Kalman filter, and is particularly useful for highly non-linear models with uncertain initial states \citep[see][for a tutorial]{mandel09}.

\begin{figure}[!t]
\centering
\noindent\includegraphics[width=\textwidth]{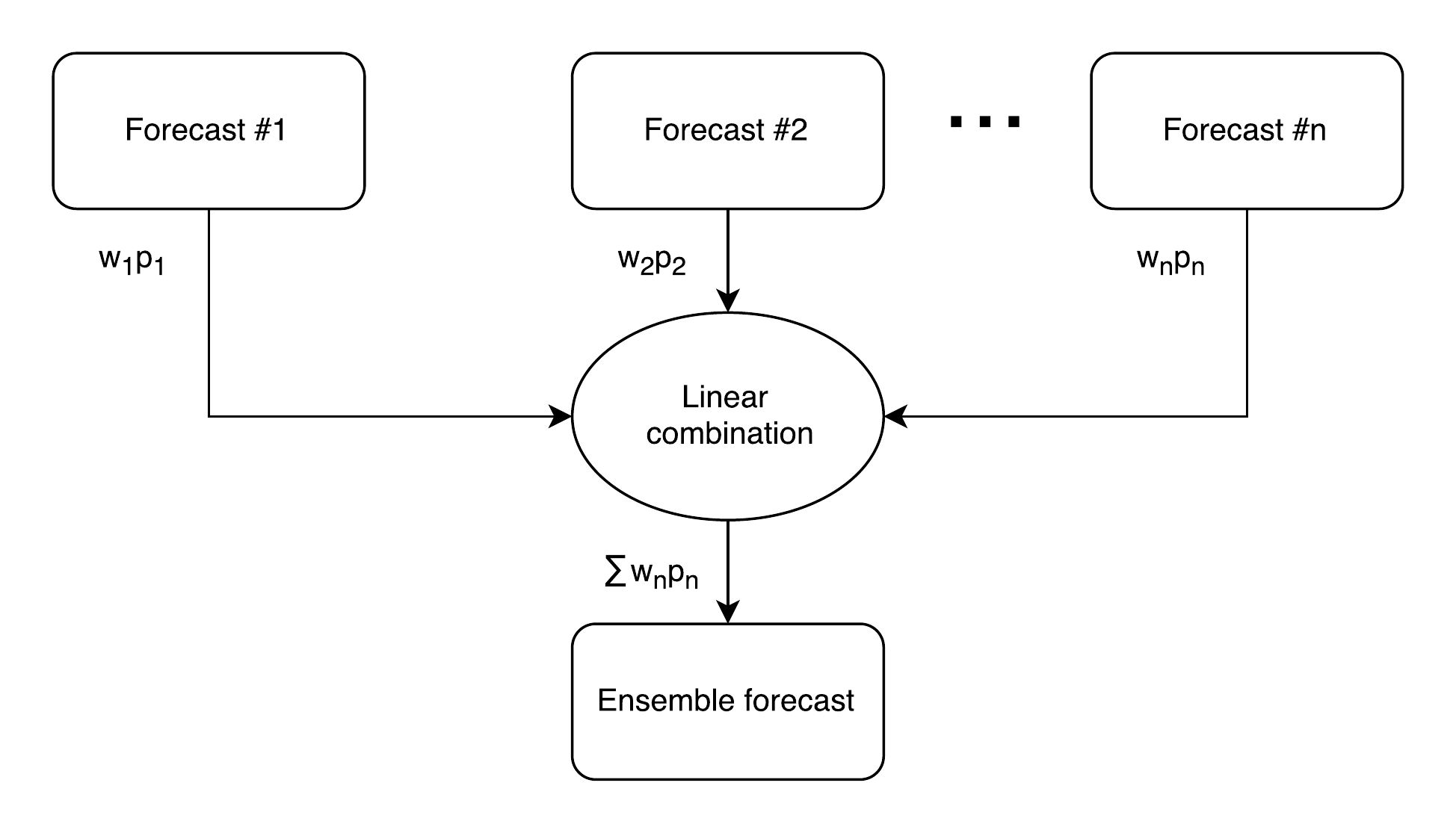}
\caption[Example of an ensemble system.]{Simple example of an ensemble of $n$ probabilistic forecasts, where $w_{n}$ denotes weights for each of the forecast probabilities, $p_{n}$, in the linear combination.}
\label{fig:ensemble}
\end{figure}

Whatever the approach, the results of these sets of forecasts can then also be used in various ways. Figure~\ref{fig:ensemble} outlines a basic example of this for probabilistic forecasts. Here, multiple forecast probabilities from different methods are linearly combined to create a best ensemble forecast probability. The resulting ensemble probability when combining $n$ forecasts is as follows,
\begin{equation}
p_{ens} = \sum_{n} \omega_{n}p_{n} ,
\end{equation}
where $p$ denotes a forecast probability value, and $w$ is some kind of weighting value, often such that the sum off all weights equals 1 \citep{zhou12}. While non-linear weighting schemes are used more widely in basic research to improve the accuracy of input models, operational forecasting methods tend to use equal-weighting schemes or simple ensemble averages \citep{genre13}. However weighting schemes based on performance metrics can be used in order to enable tailoring of forecasts based on end-user needs \citep{casanova09}, e.g., for a categorical `yes/no' forecast where an end user would like to mitigate false alarms, the weights might be determined based on performance of the false alarm ratio, (false alarms)/(hits + false alarms).

The input `forecasts' shown in Figure~\ref{fig:ensemble} could be a number of things other than probabilities depending on the sector of study, such as different models or perturbed initial conditions. Ensembles are widely used in data analytics in the private sector and beyond \citep[see e.g.,][for discussion on the renewable energy sector]{ren15}. In data science machine learning techniques such as random forests are popular \citep{breiman01}, as are combining different techniques into one predictive model in order to decrease the bias (boosting) or variance (bagging), or improve the predictive force \citep[stacking; see][for a detailed description of techniques]{dietterich00}. Ensembles have become the strategy of choice for participants of machine learning competitions such as Kaggle (\hbox{https://www.kaggle.com}), where many of the current top-performing solutions are based on ensemble methods \citep{puurula14, zou17}.

Numerical weather prediction has progressed substantially in recent decades due to the use of ensemble methods to deal with the issue of sensitivity of weather models to small initial condition changes \citep{orrell01}. For example, \citet{goerss00} found that using an ensemble forecast for the 1995-6 Atlantic hurricane season improved 48-hour tropical cyclone forecast errors by 20\% compared to the best of the individual models. Ensemble techniques have been more recently tested for probabilistic flood forecasting, for example Table 2 in \citet{jaun09} demonstrated with ranked probability skill scores (widely used in meteorology, it compares the forecast to a climatological reference) that an ensemble prediction approach performed better than deterministic alternatives for medium-range hydrological forecasts \citep[see also review by][and references therein]{cloke09}. Ensembles are nowadays assumed as standard for operations in meteorology \citep{bauer15}, to such an extent that the World Meteorological Organisation issues standardised guidelines for ensemble prediction systems \citep{wmo12}. Models often need to have an in-built capability to run in `ensemble mode' before being approved for operational use (e.g., being able to run with perturbed parameters, or its output being in a standard format so it can be combined with other model outputs). In an environment where simple, robust models are favoured over complex, but potentially more physically accurate models, ensembles allow many `weak' models to be combined to produce a `strong' model that rivals the current state-of-the-art research efforts \citep{richardson06, lin17}.  

\begin{figure}[!t]
\centering
\noindent\includegraphics[width=\textwidth]{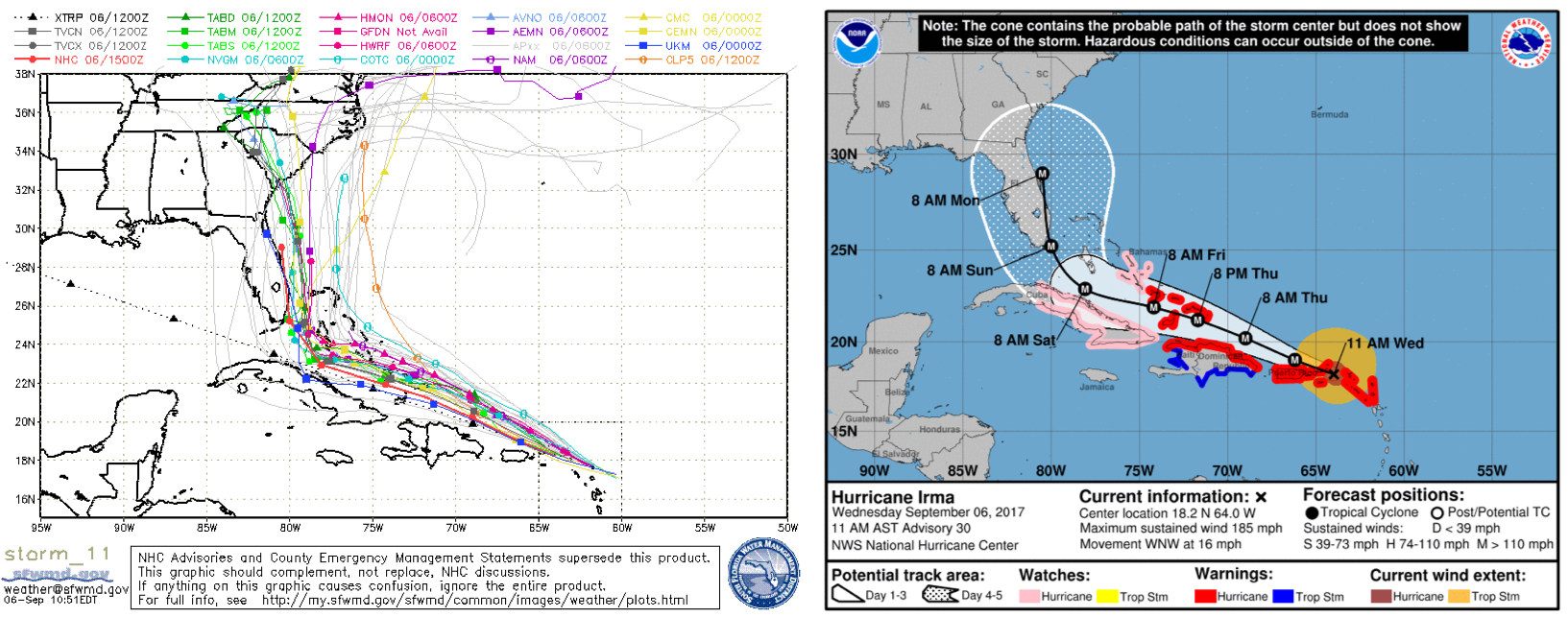}
\caption[Graphics of hurricane forecasts for Hurricane Irma in September 2017]{Graphics showing forecasts for Hurricane Irma on 6 September 2017. Spaghetti lines denote ensemble model runs from the South Florida Water Management District (https://www.sfwmd.gov/weather-radar/hurricane-model-plots) in the graphic on the left, while the cone of uncertainty from the US National Hurricane Center (https://www.nhc.noaa.gov) is shown on the right. Note that circles in the cone track are calculated from five years of historical forecast errors.}
\label{fig:hurricane}
\end{figure}

Grand ensembles are also widely used in weather and climate prediction, an ensemble of ensembles with at least two nested ensembles. For example, The International Grand Global Ensemble \citep{bougeault10} consists of ensemble forecast data from ten global numerical weather prediction centres, created to accelerate accuracy improvements of 1-day to 2-week high-impact forecasts. Grand ensembles are particularly helpful for computationally intensive research efforts such as the \hbox{climateprediction.net} project aiming to reduce uncertainties in climate modeling using volunteer computing resources \citep{allen99}. Perhaps the most widely recognisable example of ensembles to the general public lies in visualisations of hurricane forecasts \citep{padilla17}; see Figure~\ref{fig:hurricane} for examples of popular graphics, including displaying spaghetti lines showing different ensemble runs and cones of uncertainty. However properly communicating what exactly these ensembles show to ensure the public interprets the danger correctly has proven to be difficult even after decades of use. This continues to be a wider issue in weather prediction in general \citep[see e.g,][for discussion related to flood forecasts]{demeritt10}, and is an important lesson to remember in future space weather forecasting efforts.

\section{Ensembles for Space Weather}
\label{spwx_ensembles}

In recent years ensembles have become increasingly popular with space weather researchers as more adopt methods used in terrestrial weather forecasting and research \citep{henley17}. \citet{knipp16} highlighted some research works that have used ensembles to enhance models or forecasts. For example, \citet{guerra15} created a four-member multi-model ensemble for major solar flare prediction, finding an improved forecast output compared to any one single model. \citet{owens14} studied methods to improve multi-day forecasts of solar-wind-driven magnetospheric activity using a downscaling scheme, while introducing a cost-loss analysis explored again by \citet{owens17}. More recently, \citet{schunk16} created an ionosphere-thermosphere-electrodynamics multi-model ensemble prediction system based on seven physics-based data assimilation models (some using Kalman filters). The AGU 2017 frontier science session also highlighted new efforts, including the implementation of EnKF data assimilation in a solar dynamo model \citep{dikpati17}, perturbed input ensemble methods for advanced magnetospheric modeling \citep{morley17} as well as an update to the \citet{guerra15} work for operational use \citep{murray17a}.

Although ensemble methods are increasingly being used by space weather researchers, much of this research has yet to be transitioned into operations. Many forecasting centers use simple single-run, deterministic predictions, for example the ionospheric D-Region Absorption Prediction model, auroral OVATION model, and Space Weather Modeling Framework geospace model run by US NOAA Space Weather Prediction Center (SWPC) and others worldwide do not currently run in ensemble mode (see \hbox{https://www.swpc.noaa.gov/models} for model descriptions and references). Prediction of coronal mass ejection (CME) arrival time at Earth is the major exception, with SWPC and the UK's Met Office both having adapted their heliospheric propagation models to include ensembles \citep[see, e.g.,][]{cash15, pizzo15}. Some variation of the Wang-Sheeley-Arge-ENLIL model \citep{arge00, odstrcil03} is used, creating an ensemble via multiple runs of the same model but perturbing the initial CME input conditions, namely CME speed, location, and angular extent. This allows the multiple runs to provide a range of CME arrival times at Earth and other points of interest in the solar system. \citet{harrison17} recently highlighted the importance of correct input observations in operational modeling, in the context of using heliospheric imagery to prune ENLIL model ensemble forecasts.

It is worth re-iterating that an operational ensemble will be quite different to those resulting from research-focused projects, with operations generally using more simplistic robust methods. Following meteorology's example, ensemble averages have slowly started to become more widely available for space weather operational forecasting use. The NASA Community Coordinated Modeling Center CME and solar flare scoreboards (\hbox{https://ccmc.gsfc.nasa.gov/challenges}) are an excellent example of this, showing real-time results of multiple operational center forecasts from around the world as well as an average of all of them. Whilst not necessarily the best performing output, an ensemble average provides forecasters a reliable `first-guess' before more complex model runs or observational data are available, without having to decide on which of the many models results to trust.


\section{Forecast Uncertainty}

While ensemble approaches are beginning to be used in space weather operations, one area that is still consistently overlooked is forecast uncertainty. For example, solar flare forecasts are currently provided to end-users as-is, e.g., a statement will be issued that there is a 10\% chance of flares occurring in the next 24 hours \citep{murray17}. It is rare to encounter error bars included with these forecasts, e.g., $10 \pm 1\%$. It has proven to be a difficult task for many areas of space weather forecasting to accurately include such values considering the many inaccuracies involved in obtaining observations let alone within the models themselves \citep{tsagouri13}. 

\begin{figure}[!t]
\centering
\noindent\includegraphics[width=\textwidth]{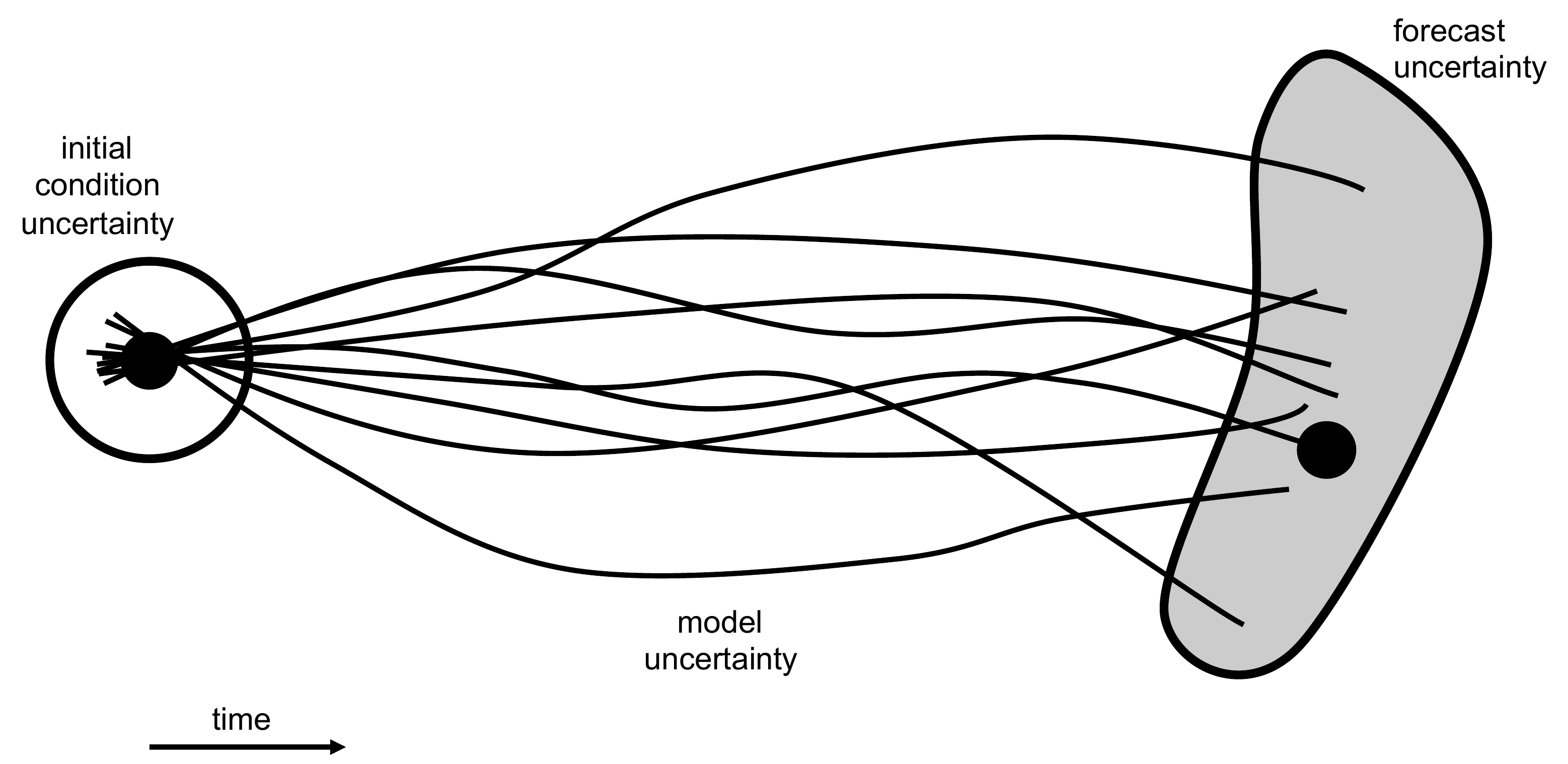}
\caption[Schematic of an ensemble forecast using initial condition uncertainties.]{Schematic of an ensemble forecast using initial condition uncertainties (not an actual forecast). The lines represent trajectories of the individual forecasts that diverge from each other owing to uncertainties in the initial conditions. The grey envelope highlights the range of states sampled by the model predictions. Note that the black circle represents the single deterministic forecast run.}
\label{fig:uncertainty}
\end{figure}

One of the main strengths of ensemble forecasting systems is their ability to represent the uncertainty which is inherent to any forecast. \citet{slingo11} discuss how probability forecasts have come to dominate weather and climate prediction and how the seminal work of \citet{lorenz69} in chaos theory has influenced how uncertainties are calculated in this field. Figure~\ref{fig:uncertainty} shows an ensemble forecast inspired by \citeauthor{slingo11}\textit{'s} example, which is based on perturbations of initial atmospheric state variable conditions. Since the Earth's atmosphere is non-linear, the perturbations are amplified by chaotic processes and each forecast diverges from the others. Here the forecast uncertainty arises from both the initial condition and model uncertainties, with the spread of a resulting ensemble forecast indicating a level of confidence in the prediction. Ensembles are useful to quantify model uncertainty and add errors bars to issued forecasts, giving the end-user a much clearer picture of the real value of the forecast. It is worth noting that for space weather, and particularly heliophysics modeling, input data uncertainty may also need to be taken into account depending on the ensemble method being used \citep[see][for a more detailed breakdown of various uncertainty types]{walker03}.

There has been some progress by space weather researchers using ensemble methods for uncertainty estimation. For example, \citet{camporeale16} used ensembles to propagate uncertainties in radiation belt simulations, perturbing the input parameters of the simulations with a collocation method. \citet{chartier16} assimilated total electron content observations into a parameter-adjusting thermosphere-ionosphere model ensemble using an EnKF.  \citet{chen18} used a Monte Carlo approach to assess the impact of input data on a radiation belt model uncertainty. \citet{dumbo18} derived arrival prediction uncertainties in CME propagation using a perturbed-initial-condition ensemble approach with a drag-based model. However estimates of uncertainty and error have yet to appear widely in operational products. Some of the newer complex models in operation at SWPC do include real-time estimates, such as the North American Total Electron Content, a Kalman filter data assimilation model (\hbox{https://www.swpc.noaa.gov/products/north-american-total-electron-content}). On the other hand, none of the operational space weather models mentioned in Section~\ref{spwx_ensembles} currently provide errors alongside their results except for the ensemble version of the ENLIL model, which includes a spread of CME arrival times \citep{mays15}.

\section{Future Efforts}
Ensemble methods are certainly useful for providing an improved forecast as well as an estimate of forecast uncertainty, and uncertainties will be utilised more frequently in future iterations of space weather predictions issued to end-users. Nevertheless, the continuing emergence of ensemble techniques in many parts of the space weather system is encouraging. Further basic research and development with ensembles, particularly in areas such as solar eruptive forecasting, may continue to provide useful insights that can be transitioned to enhanced space weather operations. This may ultimately aid space weather forecasts to begin to rival meteorological efforts in terms of accuracy. The power of a simple ensemble average is not to be underestimated in the meantime.


\acknowledgments
S.A.M. is supported by the Irish Research Council Post-doctoral Fellowship Programme and the Air Force Office of Scientific Research award number FA9550-17-1-039. No data was generated or used in this manuscript. The author would like to thank the anonymous referees and editor for constructive suggestions to improve the manuscript.


\begin{thebibliography}{51}
\providecommand{\natexlab}[1]{#1}
\expandafter\ifx\csname urlstyle\endcsname\relax
  \providecommand{\doi}[1]{doi:\discretionary{}{}{}#1}\else
  \providecommand{\doi}{doi:\discretionary{}{}{}\begingroup
  \urlstyle{rm}\Url}\fi

\bibitem[{\textit{{Allen}}(1999)}]{allen99}
{Allen}, M. (1999), {Do-it-yourself climate prediction}, \textit{Nature},
  \textit{401}, 642, \doi{10.1038/44266}.

\bibitem[{\textit{{Arge} and {Pizzo}}(2000)}]{arge00}
{Arge}, C.~N., and V.~J. {Pizzo} (2000), {Improvement in the prediction of
  solar wind conditions using near-real time solar magnetic field updates},
  \textit{Journal of Geophysical Research}, \textit{105}, 10,465--10,480,
  \doi{10.1029/1999JA000262}.

\bibitem[{\textit{{Barnes} et~al.}(2016)\textit{{Barnes}, {Leka}, {Schrijver},
  {Colak}, {Qahwaji}, {Ashamari}, {Yuan}, {Zhang}, {McAteer}, {Bloomfield},
  {Higgins}, {Gallagher}, {Falconer}, {Georgoulis}, {Wheatland}, {Balch},
  {Dunn}, and {Wagner}}}]{barnes16}
{Barnes}, G., K.~D. {Leka}, C.~J. {Schrijver}, T.~{Colak}, R.~{Qahwaji}, O.~W.
  {Ashamari}, Y.~{Yuan}, J.~{Zhang}, R.~T.~J. {McAteer}, D.~S. {Bloomfield},
  P.~A. {Higgins}, P.~T. {Gallagher}, D.~A. {Falconer}, M.~K. {Georgoulis},
  M.~S. {Wheatland}, C.~{Balch}, T.~{Dunn}, and E.~L. {Wagner} (2016), {A
  Comparison of Flare Forecasting Methods. I. Results from the All-Clear
  Workshop}, \textit{Astrophysical Journal}, \textit{829}, 89,
  \doi{10.3847/0004-637X/829/2/89}.

\bibitem[{\textit{{Bauer} et~al.}(2015)\textit{{Bauer}, {Thorpe}, and
  {Brunet}}}]{bauer15}
{Bauer}, P., A.~{Thorpe}, and G.~{Brunet} (2015), {The quiet revolution of
  numerical weather prediction}, \textit{Nature}, \textit{525}, 47--55,
  \doi{10.1038/nature14956}.

\bibitem[{\textit{{Bobra}}(2017)}]{bobra17}
{Bobra}, M. (2017), {Predicting Solar Activity Using Machine-Learning Methods},
  \textit{AGU Fall Meeting Abstracts}.

\bibitem[{\textit{Bougeault et~al.}(2010)\textit{Bougeault, Toth, Bishop,
  Brown, Burridge, Chen, Ebert, Fuentes, Hamill, Mylne, Nicolau, Paccagnella,
  Park, Parsons, Raoult, Schuster, Dias, Swinbank, Takeuchi, Tennant, Wilson,
  and Worley}}]{bougeault10}
Bougeault, P., Z.~Toth, C.~Bishop, B.~Brown, D.~Burridge, D.~H. Chen, B.~Ebert,
  M.~Fuentes, T.~M. Hamill, K.~Mylne, J.~Nicolau, T.~Paccagnella, Y.-Y. Park,
  D.~Parsons, B.~Raoult, D.~Schuster, P.~S. Dias, R.~Swinbank, Y.~Takeuchi,
  W.~Tennant, L.~Wilson, and S.~Worley (2010), The thorpex interactive grand
  global ensemble, \textit{Bulletin of the American Meteorological Society},
  \textit{91}(8), 1059--1072, \doi{10.1175/2010BAMS2853.1}.

\bibitem[{\textit{Breiman}(2001)}]{breiman01}
Breiman, L. (2001), Random forests, \textit{Machine Learning}, \textit{45}(1),
  5--32, \doi{10.1023/A:1010933404324}.

\bibitem[{\textit{{Camporeale} et~al.}(2016)\textit{{Camporeale}, {Shprits},
  {Chandorkar}, {Drozdov}, and {Wing}}}]{camporeale16}
{Camporeale}, E., Y.~{Shprits}, M.~{Chandorkar}, A.~{Drozdov}, and S.~{Wing}
  (2016), {On the propagation of uncertainties in radiation belt simulations},
  \textit{Space Weather}, \textit{14}, 982--992, \doi{10.1002/2016SW001494}.

\bibitem[{\textit{Casanova and Ahrens}(2009)}]{casanova09}
Casanova, S., and B.~Ahrens (2009), On the weighting of multimodel ensembles in
  seasonal and short-range weather forecasting, \textit{Monthly Weather
  Review}, \textit{137}(11), 3811--3822, \doi{10.1175/2009MWR2893.1}.

\bibitem[{\textit{{Cash} et~al.}(2015)\textit{{Cash}, {Biesecker}, {Pizzo}, {de
  Koning}, {Millward}, {Arge}, {Henney}, and {Odstrcil}}}]{cash15}
{Cash}, M.~D., D.~A. {Biesecker}, V.~{Pizzo}, C.~A. {de Koning}, G.~{Millward},
  C.~N. {Arge}, C.~J. {Henney}, and D.~{Odstrcil} (2015), {Ensemble Modeling of
  the 23 July 2012 Coronal Mass Ejection}, \textit{Space Weather}, \textit{13},
  611--625, \doi{10.1002/2015SW001232}.

\bibitem[{\textit{Chartier et~al.}(2016)\textit{Chartier, Matsuo, Anderson,
  Collins, Hoar, Lu, Mitchell, Coster, Paxton, and Bust}}]{chartier16}
Chartier, A.~T., T.~Matsuo, J.~L. Anderson, N.~Collins, T.~J. Hoar, G.~Lu,
  C.~N. Mitchell, A.~J. Coster, L.~J. Paxton, and G.~S. Bust (2016),
  Ionospheric data assimilation and forecasting during storms, \textit{Journal
  of Geophysical Research: Space Physics}, \textit{121}(1), 764--778,
  \doi{10.1002/2014JA020799}, 2014JA020799.

\bibitem[{\textit{{Chen} et~al.}(2018)\textit{{Chen}, {O'Brien}, {Lemon}, and
  {Guild}}}]{chen18}
{Chen}, M.~W., T.~P. {O'Brien}, C.~L. {Lemon}, and T.~B. {Guild} (2018),
  {Effects of Uncertainties in Electric Field Boundary Conditions for Ring
  Current Simulations}, \textit{Journal of Geophysical Research (Space
  Physics)}, \textit{123}, 638--652, \doi{10.1002/2017JA024496}.

\bibitem[{\textit{{Cloke} and {Pappenberger}}(2009)}]{cloke09}
{Cloke}, H.~L., and F.~{Pappenberger} (2009), {Ensemble flood forecasting: A
  review}, \textit{Journal of Hydrology}, \textit{375}, 613--626,
  \doi{10.1016/j.jhydrol.2009.06.005}.

\bibitem[{\textit{Demeritt et~al.}(2010)\textit{Demeritt, Nobert, Cloke, and
  Pappenberger}}]{demeritt10}
Demeritt, D., S.~Nobert, H.~Cloke, and F.~Pappenberger (2010), Challenges in
  communicating and using ensembles in operational flood forecasting,
  \textit{Meteorological Applications}, \textit{17}(2), 209--222,
  \doi{10.1002/met.194}.

\bibitem[{\textit{Dietterich}(2000)}]{dietterich00}
Dietterich, T.~G. (2000), Ensemble methods in machine learning, in
  \textit{Multiple Classifier Systems}, pp. 1--15, Springer Berlin Heidelberg,
  Berlin, Heidelberg.

\bibitem[{\textit{{Dikpati}}(2017)}]{dikpati17}
{Dikpati}, M. (2017), {Ensemble Kalman Filter Data Assimilation in a Solar
  Dynamo Model}, \textit{AGU Fall Meeting Abstracts}.

\bibitem[{\textit{{Dumbovi{\'c}} et~al.}(2018)\textit{{Dumbovi{\'c}}, {{\v
  C}alogovi{\'c}}, {Vr{\v s}nak}, {Temmer}, {Mays}, {Veronig}, and
  {Piantschitsch}}}]{dumbo18}
{Dumbovi{\'c}}, M., J.~{{\v C}alogovi{\'c}}, B.~{Vr{\v s}nak}, M.~{Temmer},
  M.~L. {Mays}, A.~{Veronig}, and I.~{Piantschitsch} (2018), {The Drag-based
  Ensemble Model (DBEM) for Coronal Mass Ejection Propagation}, \textit{The
  Astrophysical Journal}, \textit{854}, 180, \doi{10.3847/1538-4357/aaaa66}.

\bibitem[{\textit{Genre et~al.}(2013)\textit{Genre, Kenny, Meyler, and
  Timmermann}}]{genre13}
Genre, V., G.~Kenny, A.~Meyler, and A.~Timmermann (2013), Combining expert
  forecasts: Can anything beat the simple average?, \textit{International
  Journal of Forecasting}, \textit{29}(1), 108 -- 121,
  \doi{https://doi.org/10.1016/j.ijforecast.2012.06.004}.

\bibitem[{\textit{{Goerss}}(2000)}]{goerss00}
{Goerss}, J.~S. (2000), {Tropical Cyclone Track Forecasts Using an Ensemble of
  Dynamical Models}, \textit{Monthly Weather Review}, \textit{128}, 1187,
  \doi{10.1175/1520-0493(2000)128<1187:TCTFUA>2.0.CO;2}.

\bibitem[{\textit{{Guerra} et~al.}(2015)\textit{{Guerra}, {Pulkkinen}, and
  {Uritsky}}}]{guerra15}
{Guerra}, J.~A., A.~{Pulkkinen}, and V.~M. {Uritsky} (2015), {Ensemble
  forecasting of major solar flares: First results}, \textit{Space Weather},
  \textit{13}, 626--642, \doi{10.1002/2015SW001195}.

\bibitem[{\textit{Harrison et~al.}(2017)\textit{Harrison, Davies, Biesecker,
  and Gibbs}}]{harrison17}
Harrison, R.~A., J.~A. Davies, D.~Biesecker, and M.~Gibbs (2017), The
  application of heliospheric imaging to space weather operations: Lessons
  learned from published studies, \textit{Space Weather}, \textit{15}(8),
  985--1003, \doi{10.1002/2017SW001633}, 2017SW001633.

\bibitem[{\textit{Henley and Pope}(2017)}]{henley17}
Henley, E.~M., and E.~C.~D. Pope (2017), Cost-loss analysis of ensemble solar
  wind forecasting: Space weather use of terrestrial weather tools,
  \textit{Space Weather}, \textit{15}(12), 1562--1566,
  \doi{10.1002/2017SW001758}, 2017SW001758.

\bibitem[{\textit{{Jaun} and {Ahrens}}(2009)}]{jaun09}
{Jaun}, S., and B.~{Ahrens} (2009), {Evaluation of a probabilistic
  hydrometeorological forecast system}, \textit{Hydrology and Earth System
  Sciences Discussions}, \textit{6}, 1843--1877.

\bibitem[{\textit{{Johnson} and {Wing}}(2017)}]{johnson17}
{Johnson}, J., and S.~{Wing} (2017), {Tools for Detecting Causality in Space
  Systems}, \textit{AGU Fall Meeting Abstracts}.

\bibitem[{\textit{{Knipp}}(2016)}]{knipp16}
{Knipp}, D.~J. (2016), {Advances in Space Weather Ensemble Forecasting},
  \textit{Space Weather}, \textit{14}, 52--53, \doi{10.1002/2016SW001366}.

\bibitem[{\textit{Koskinen et~al.}(2017)\textit{Koskinen, Baker, Balogh,
  Gombosi, Veronig, and von Steiger}}]{koskinen17}
Koskinen, H. E.~J., D.~N. Baker, A.~Balogh, T.~Gombosi, A.~Veronig, and R.~von
  Steiger (2017), Achievements and challenges in the science of space weather,
  \textit{Space Science Reviews}, \textit{212}(3), 1137--1157,
  \doi{10.1007/s11214-017-0390-4}.

\bibitem[{\textit{{Lin} et~al.}(2017)\textit{{Lin}, {Bender}, {Harris}, and
  {Hazelton}}}]{lin17}
{Lin}, S.~J., M.~{Bender}, L.~{Harris}, and A.~{Hazelton} (2017), {Performance
  of the FV3-powered Next Generation Global Prediction System for Harvey and
  Irma, and a vision for a ``beyond weather timescale'' prediction system for
  long-range hurricane track and intensity predictions}, \textit{AGU Fall
  Meeting Abstracts}.

\bibitem[{\textit{{Lorenz}}(1969)}]{lorenz69}
{Lorenz}, E.~N. (1969), {The predictability of a flow which possesses many
  scales of motion}, \textit{Tellus}, \textit{21}, 289--307.

\bibitem[{\textit{{Mandel}}(2009)}]{mandel09}
{Mandel}, J. (2009), {A Brief Tutorial on the Ensemble Kalman Filter},
  \textit{ArXiv e-prints}.

\bibitem[{\textit{{Mays} et~al.}(2015)\textit{{Mays}, {Taktakishvili},
  {Pulkkinen}, {MacNeice}, {Rast{\"a}tter}, {Odstrcil}, {Jian}, {Richardson},
  {LaSota}, {Zheng}, and {Kuznetsova}}}]{mays15}
{Mays}, M.~L., A.~{Taktakishvili}, A.~{Pulkkinen}, P.~J. {MacNeice},
  L.~{Rast{\"a}tter}, D.~{Odstrcil}, L.~K. {Jian}, I.~G. {Richardson}, J.~A.
  {LaSota}, Y.~{Zheng}, and M.~M. {Kuznetsova} (2015), {Ensemble Modeling of
  CMEs Using the WSA-ENLIL+Cone Model}, \textit{Solar Physics}, \textit{290},
  1775--1814, \doi{10.1007/s11207-015-0692-1}.

\bibitem[{\textit{McGranaghan et~al.}(2017)\textit{McGranaghan, Bhatt, Matsuo,
  Mannucci, Semeter, and Datta-Barua}}]{mcgranaghan17}
McGranaghan, R.~M., A.~Bhatt, T.~Matsuo, A.~J. Mannucci, J.~L. Semeter, and
  S.~Datta-Barua (2017), Ushering in a new frontier in geospace through data
  science, \textit{Journal of Geophysical Research: Space Physics},
  \textit{122}(12), 12,586--12,590, \doi{10.1002/2017JA024835}, 2017JA024835.

\bibitem[{\textit{{Morley} et~al.}(2017)\textit{{Morley}, {Steinberg},
  {Haiducek}, {Welling}, {Hassan}, and {Weaver}}}]{morley17}
{Morley}, S., J.~T. {Steinberg}, J.~D. {Haiducek}, D.~T. {Welling},
  E.~{Hassan}, and B.~P. {Weaver} (2017), {Perturbed-input-data ensemble
  modeling of magnetospheric dynamics}, \textit{AGU Fall Meeting Abstracts}.

\bibitem[{\textit{{Murray} and {Guerra}}(2017)}]{murray17a}
{Murray}, S., and J.~A. {Guerra} (2017), {Ensemble flare forecasting: using
  numerical weather prediction techniques to improve space weather operations},
  \textit{AGU Fall Meeting Abstracts}.

\bibitem[{\textit{Murray et~al.}(2017)\textit{Murray, Bingham, Sharpe, and
  Jackson}}]{murray17}
Murray, S.~A., S.~Bingham, M.~Sharpe, and D.~R. Jackson (2017), Flare
  forecasting at the met office space weather operations centre, \textit{Space
  Weather}, \textit{15}(4), 577--588, \doi{10.1002/2016SW001579}, 2016SW001579.

\bibitem[{\textit{{Odstrcil}}(2003)}]{odstrcil03}
{Odstrcil}, D. (2003), {Modeling 3-D solar wind structure}, \textit{Advances in
  Space Research}, \textit{32}, 497--506, \doi{10.1016/S0273-1177(03)00332-6}.

\bibitem[{\textit{Orrell et~al.}(2001)\textit{Orrell, Smith, Barkmeijer, and
  Palmer}}]{orrell01}
Orrell, D., L.~Smith, J.~Barkmeijer, and T.~N. Palmer (2001), Model error in
  weather forecasting, \textit{Nonlinear Processes in Geophysics},
  \textit{8}(6), 357--371, \doi{10.5194/npg-8-357-2001}.

\bibitem[{\textit{Owens and Riley}(2017)}]{owens17}
Owens, M.~J., and P.~Riley (2017), Probabilistic solar wind forecasting using
  large ensembles of near-sun conditions with a simple one-dimensional upwind
  scheme, \textit{Space Weather}, \textit{15}(11), 1461--1474,
  \doi{10.1002/2017SW001679}, 2017SW001679.

\bibitem[{\textit{Owens et~al.}(2014)\textit{Owens, Horbury, Wicks, McGregor,
  Savani, and Xiong}}]{owens14}
Owens, M.~J., T.~S. Horbury, R.~T. Wicks, S.~L. McGregor, N.~P. Savani, and
  M.~Xiong (2014), Ensemble downscaling in coupled solar wind-magnetosphere
  modeling for space weather forecasting, \textit{Space Weather},
  \textit{12}(6), 395--405, \doi{10.1002/2014SW001064}.

\bibitem[{\textit{Padilla et~al.}(2017)\textit{Padilla, Ruginski, and
  Creem-Regehr}}]{padilla17}
Padilla, L.~M., I.~T. Ruginski, and S.~H. Creem-Regehr (2017), Effects of
  ensemble and summary displays on interpretations of geospatial uncertainty
  data, \textit{Cognitive Research: Principles and Implications},
  \textit{2}(1), 40, \doi{10.1186/s41235-017-0076-1}.

\bibitem[{\textit{{Pizzo} et~al.}(2015)\textit{{Pizzo}, {de Koning}, {Cash},
  {Millward}, {Biesecker}, {Puga}, {Codrescu}, and {Odstrcil}}}]{pizzo15}
{Pizzo}, V.~J., C.~{de Koning}, M.~{Cash}, G.~{Millward}, D.~A. {Biesecker},
  L.~{Puga}, M.~{Codrescu}, and D.~{Odstrcil} (2015), {Theoretical basis for
  operational ensemble forecasting of coronal mass ejections}, \textit{Space
  Weather}, \textit{13}, 676--697, \doi{10.1002/2015SW001221}.

\bibitem[{\textit{Puurula et~al.}(2014)\textit{Puurula, Read, and
  Bifet}}]{puurula14}
Puurula, A., J.~Read, and A.~Bifet (2014), Kaggle lshtc4 winning solution,
  \textit{CoRR}, \textit{abs/1405.0546}.

\bibitem[{\textit{Ren et~al.}(2015)\textit{Ren, Suganthan, and
  Srikanth}}]{ren15}
Ren, Y., P.~Suganthan, and N.~Srikanth (2015), Ensemble methods for wind and
  solar power forecasting: A state-of-the-art review, \textit{Renewable and
  Sustainable Energy Reviews}, \textit{50}, 82 -- 91,
  \doi{https://doi.org/10.1016/j.rser.2015.04.081}.

\bibitem[{\textit{Richardson}(2006)}]{richardson06}
Richardson, D.~S. (2006), Skill and relative economic value of the ecmwf
  ensemble prediction system, \textit{Quarterly Journal of the Royal
  Meteorological Society}, \textit{126}(563), 649--667,
  \doi{10.1002/qj.49712656313}.

\bibitem[{\textit{{Schunk} et~al.}(2016)\textit{{Schunk}, {Scherliess},
  {Eccles}, {Gardner}, {Sojka}, {Zhu}, {Pi}, {Mannucci}, {Butala}, {Wilson},
  {Komjathy}, {Wang}, and {Rosen}}}]{schunk16}
{Schunk}, R.~W., L.~{Scherliess}, V.~{Eccles}, L.~C. {Gardner}, J.~J. {Sojka},
  L.~{Zhu}, X.~{Pi}, A.~J. {Mannucci}, M.~{Butala}, B.~D. {Wilson},
  A.~{Komjathy}, C.~{Wang}, and G.~{Rosen} (2016), {Space weather forecasting
  with a Multimodel Ensemble Prediction System (MEPS)}, \textit{Radio Science},
  \textit{51}, 1157--1165, \doi{10.1002/2015RS005888}.

\bibitem[{\textit{Sharpe and Murray}(2017)}]{sharpe17}
Sharpe, M.~A., and S.~A. Murray (2017), Verification of space weather forecasts
  issued by the met office space weather operations centre, \textit{Space
  Weather}, \textit{15}(10), 1383--1395, \doi{10.1002/2017SW001683},
  2017SW001683.

\bibitem[{\textit{Slingo and Palmer}(2011)}]{slingo11}
Slingo, J., and T.~Palmer (2011), Uncertainty in weather and climate
  prediction, \textit{Philosophical Transactions of the Royal Society of London
  A: Mathematical, Physical and Engineering Sciences}, \textit{369}(1956),
  4751--4767, \doi{10.1098/rsta.2011.0161}.

\bibitem[{\textit{{Tsagouri} et~al.}(2013)\textit{{Tsagouri}, {Belehaki},
  {Bergeot}, {Cid}, {Delouille}, {Egorova}, {Jakowski}, {Kutiev}, {Mikhailov},
  {Núñez}, {Pietrella}, {Potapov}, {Qahwaji}, {Tulunay}, {Velinov}, and
  {Viljanen}}}]{tsagouri13}
{Tsagouri}, I., A.~{Belehaki}, N.~{Bergeot}, C.~{Cid}, V.~{Delouille},
  T.~{Egorova}, N.~{Jakowski}, I.~{Kutiev}, A.~{Mikhailov}, M.~{Núñez},
  M.~{Pietrella}, A.~{Potapov}, R.~{Qahwaji}, Y.~{Tulunay}, P.~{Velinov}, and
  A.~{Viljanen} (2013), Progress in space weather modeling in an operational
  environment, \textit{J. Space Weather Space Clim.}, \textit{3}, A17,
  \doi{10.1051/swsc/2013037}.

\bibitem[{\textit{Walker et~al.}(2003)\textit{Walker, Harremoës, Rotmans,
  van~der Sluijs, van Asselt, Janssen, and von Krauss}}]{walker03}
Walker, W., P.~Harremoës, J.~Rotmans, J.~van~der Sluijs, M.~van Asselt,
  P.~Janssen, and M.~K. von Krauss (2003), Defining uncertainty: A conceptual
  basis for uncertainty management in model-based decision support,
  \textit{Integrated Assessment}, \textit{4}(1), 5--17,
  \doi{10.1076/iaij.4.1.5.16466}.

\bibitem[{\textit{WMO}(2012)}]{wmo12}
WMO (2012), \textit{World Meteorological Organisation Guidelines on Ensemble
  Prediction Systems and Forecasting}, WMO-No. 1091.

\bibitem[{\textit{Zhou}(2012)}]{zhou12}
Zhou, Z.-H. (2012), \textit{Ensemble Methods: Foundations and Algorithms}, 1st
  ed., Chapman \& Hall/CRC.

\bibitem[{\textit{Zou et~al.}(2017)\textit{Zou, Xu, and Li}}]{zou17}
Zou, H., K.~Xu, and J.~Li (2017), The youtube-8m kaggle competition: Challenges
  and methods, \textit{CoRR}, \textit{abs/1706.09274}.

\end{thebibliography}

\end{document}